\documentclass[a4paper, amsfonts, amssymb, amsmath, reprint, showkeys, nofootinbib, twoside]{revtex4-1}
\usepackage[english]{babel}
\usepackage[utf8]{inputenc}
\usepackage[colorinlistoftodos, color=green!40, prependcaption]{todonotes}
\usepackage{amsthm}
\usepackage{mathtools}
\usepackage{physics}
\usepackage{xcolor}
\usepackage{graphicx}
\usepackage[left=23mm,right=13mm,top=35mm,columnsep=15pt]{geometry} 
\usepackage{adjustbox}
\usepackage{placeins}
\usepackage[T1]{fontenc}
\usepackage{lipsum}
\usepackage{csquotes}
\usepackage[pdftex, pdftitle={Article}, pdfauthor={Author}]{hyperref} % For hyperlinks in the PDF
\begin{document}

\title{A detection module to increase the spatial bandwidth product of a microscope}

\author{Charlie Keruszan}
    \affiliation{LP2N, Laboratoire Photonique Numéerique et Nanosciences, Univ. Bordeaux, F-33400 Talence, France}
    \affiliation{Institut d'Optique Graduate School $\&$ CNRS UMR 5298, F-33400 Talence, France}

\author{Aymerick Bazin}
    \affiliation{LP2N, Laboratoire Photonique Numéerique et Nanosciences, Univ. Bordeaux, F-33400 Talence, France}
    \affiliation{Institut d'Optique Graduate School $\&$ CNRS UMR 5298, F-33400 Talence, France}

\author{Amaury Badon}
    \email[Correspondence email address: ]{amaury.badon@cnrs.fr}% Your name
    \affiliation{LP2N, Laboratoire Photonique Numéerique et Nanosciences, Univ. Bordeaux, F-33400 Talence, France}
    \affiliation{Institut d'Optique Graduate School $\&$ CNRS UMR 5298, F-33400 Talence, France}
    
\date{\today} % Leave empty to omit a date

\begin{abstract}
Capturing biological specimens at large scales with sub-micron resolution is crucial for biomedical research, but conventional cameras often can't handle the pixel requirements. While most microscopes use motorized stages to move samples and capture images tile by tile, we propose a method that eliminates the need for sample movement. Our approach integrates a scanning mechanism within the microscope's detection unit, enabling sequential capture of sub-areas of the field of view without physically moving the sample. This "remote scanning" method works with all camera-based microscopes and we demonstrate its effectiveness in both bright-field and epifluorescence microscopy. This technique is ideal for imaging motion-sensitive samples and large biological specimens, as evidenced in this study with millimeter-scale engineered tissues.
\end{abstract}

\maketitle

%%%%%%%%%%%%%%%%%%%%%%%%%%  body  %%%%%%%%%%%%%%%%%%%%%%%%%%

Over the past few years, new imaging systems have emerged to capture subcellular spatial information across fields of view (FOV) that are orders of magnitude larger, typically ranging from $\text{mm}^2$ to $\text{cm}^2$ \cite{yang2022daxi,glaser2022hybrid}. These high-content imaging techniques have been particularly useful in the context of the study of embryonic development and in the field of tissues engineering. In the latter field, it is crucial to ensure that tissue organization closely replicates \textit{in vivo} conditions across multiple spatial scales to validate the relevance of these models. 
The simplest approach to imaging large-scale samples at diffraction-limited resolution involves moving the sample stage and acquiring images at different positions. While effective for fixed samples, such as in histopathology or tissue clearing applications, this method is slow and incompatible with alive or motion-sensitive samples. In the absence of motion, the amount of information a microscope can capture is fundamentally constrained by its space-bandwidth product (SBP) \cite{lohmann1996space}. This parameter, commonly used to characterize microscope objectives, represents the number of pixels required to image the entire FOV while fullfilling Nyquist sampling criterion \cite{park2021review}. As a reference point, a X10 objective with a 0.67 $\mu$m lateral resolution and a theoretical 2.2 × 2.2 $\text{mm}^2$  FOV has an SBP of 47 megapixels (Nikon, CFI Super Fluor, 10X NA=0.5) for a wavelength equals to 500 nm. 
In point-scanning imaging techniques, such as confocal microscopy, this SBP can be achieved by finely scanning the FOV with a diffraction-limited illumination spot. However, in camera-based microscopy, this task is more challenging, as the SBP vastly exceeds the typical number of pixels available in scientific cameras, which usually ranges from 2 to 10 megapixels. As a result, detectors become the limiting factor of the SBP of the system and a significant portion of the information transmitted by the MO is discarded.\\
Although new sensors with increasing pixel counts—up to 150 megapixels—are being developed, they remain expensive, less efficient than scientific cameras, and generate an overwhelming amount of data, much of which is unnecessary if only specific regions of the FOV require high-resolution imaging. Alternatively, recent techniques have been proposed to enhance the SBP, either in the spatial or spectral domain 
\cite{park2021review}. While these approaches, such as Fourier ptychography \cite{zheng2013wide}, camera-array imaging \cite{brady2012multiscale} and structured illumination microscopy \cite{gustafsson2000surpassing} offer impressive performances, they often require entirely specialized setups and heavy computational approaches, limiting their accessibility and widespread use. 

In this letter, we introduce a detection module compatible with any commercially available inverted microscope, enabling simultaneous observation of a sample on two distinct channels: one covering the entire FOV at moderate resolution and another capturing a smaller region at diffraction limited lateral resolution. The proposed module is shown in
Fig. 1(a) and consists of a few optical elements positioned on a 30 $\times$ 25 $\text{cm}^2$ breadboard.
\begin{figure}[ht!]
\includegraphics[width=\linewidth]{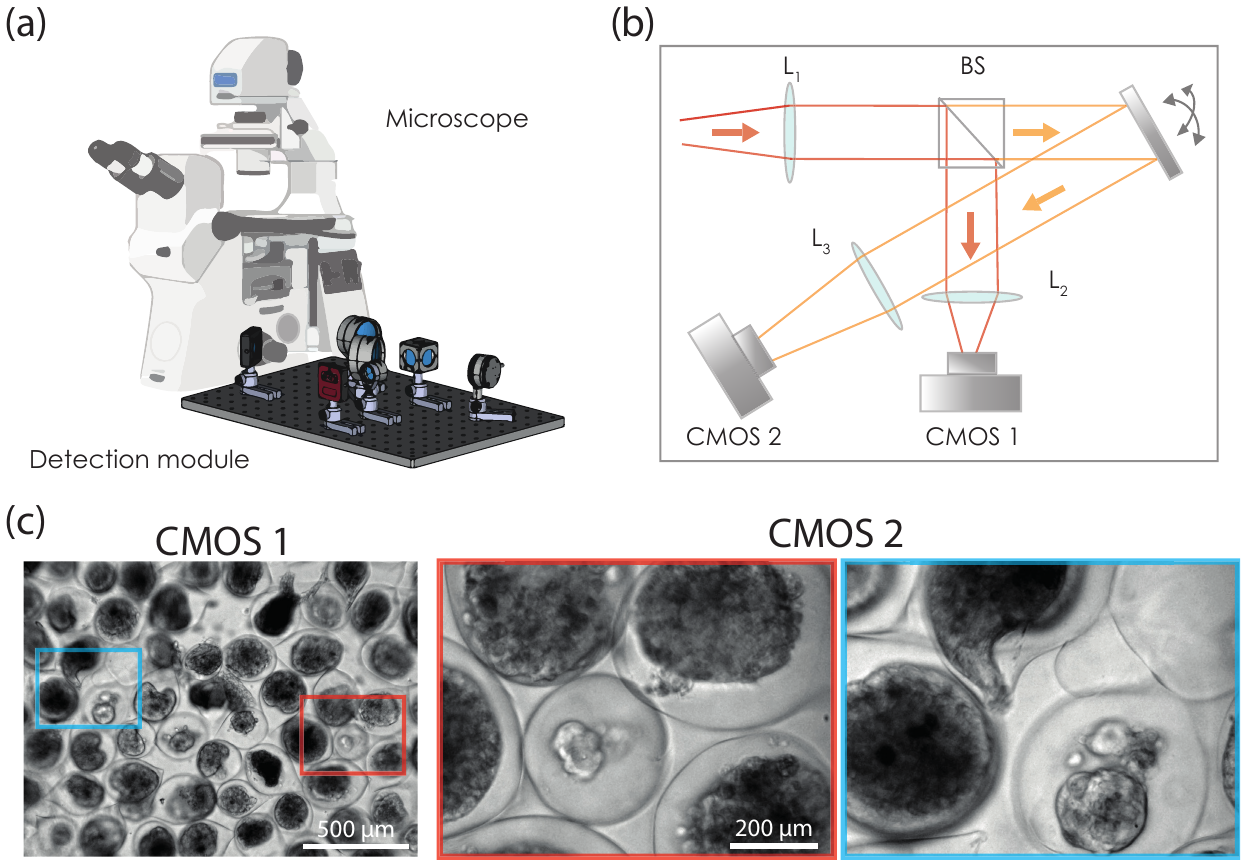}
\caption{Detection module principle. (a) 3D view of the experimental setup comprised of a commercial inversed microscope and the detection unit. (b) Schematic of the detection unit only. Light from the microscope is splitted in two by a BS. Each channel provides a different magnification which is translated into different FOV and lateral resolution. (c) Comparison of the images obtained with the first and second detection channel. Areas highlighted in red and blue on the left image correspond to the images obtained for two different angles of the steering mirror.}
\label{figure1}
\end{figure}
In detail, output light from the microscope body propagates through a lens $L_1$  (f = 150 mm) and is divided in two channels by a non-polarizing beamsplitter (BS). On the first channel, light is subsequently focused by a lens $L_2$ (f = 50 mm) onto a scientific camera ($\text{CMOS}_1$, Thorlabs Zelux).  Compared to the native image of the microscope, the afocal system made of $L_1$ and $L_2$ (magnification = 1/3) ensures that $\text{CMOS}_1$ captures a large proportion of the FOV of the MO.  By using a 10X microscope objective with a NA=0.5 (Nikon, CFI Super Fluor 10X NA=0.5), this channel provides images that cover approximately 1.8 mm x 1.35 mm with a pixel apparent size of 1.25 $\mu$m. 
In the second channel, after being transmitted by the BS, light is reflected by a motorized steering mirror (Optotune, MR-15-30) and focused by a lens $L_3$ (f=200 mm) onto a second identical scientific camera ($\text{CMOS}_2$, Thorlabs, Zelux). Under the same experimental conditions, each pixel of $\text{CMOS}_2$ maps 310 nm, resulting in a field of view of 230 $\mu$m x 173 $\mu$m. In this study, all experiments are carried out with this same 10X MO. Because the motorized 2D mirror is in the Fourier space, a tilt of the mirror is supposed to be solely translated into a shift of the image in the camera plane without introducing any image distortion.  By adjusting the shift on the motorized mirror,  any region of the FOV of the MO can be captured on $\text{CMOS}_2$.

As a first demonstration, we imaged in bright-field mode multi-cellular aggregates obtained from the encapsulation of human pluripotent stem cells in alginate shell \cite{alessandri2013cellular,cohen2023engineering}. As seen on figure \ref{figure1}c, the first detection channel provides a FOV large enough to cover several encapsulated aggregates of approximately 300 $\mu$m in diameter and gives a quick overview of the sample. Such image allows for the counting of capsules, determining whether they are empty or contain a multicellular aggregate, estimating the approximate size of the cellular aggregate, \textit{etc}.  With a much smaller FOV, the second detection channel covers only a part of a capsule but with finer details that can be used to investigate closely the structure of the capsule itself or to obtain sub-cellular details on the aggregates. In only a few milliseconds, a different tilt applied on the motorized mirror can shift the observed region of interest (ROI) on the same capsule or even on a different capsule present in the FOV of the MO. The cameras and the motorized mirror are controlled with a Python-based code with a graphical user interface that offers live visualization and adjustable acquisition parameters. This code is freely available on GitHub \cite{github_bioflab}.

\begin{figure}[ht!]
\includegraphics[width=\linewidth]{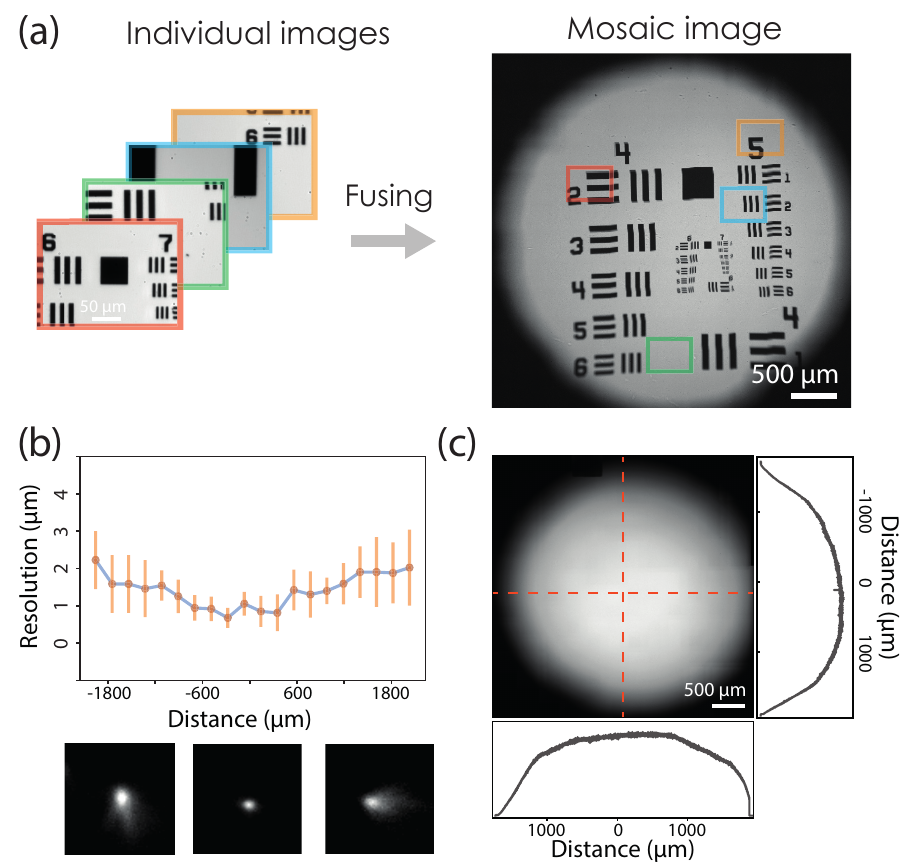}
  \caption{Optical performances and image fusing. (a) A stack of images captured for different angles of the detection unit is rearranged as a mosaic to cover the entire FOV of the MO. (b) Spatial resolution of the system depending on the location inside the microscope FOV. (c) Collection efficiency of the system measured in bright-field mode and without sample and plot profile as insets. }
\label{figure2}
\end{figure}
\begin{center}
\begin{figure*}[ht!]
\begin{center}
    \includegraphics[width=0.8\linewidth]{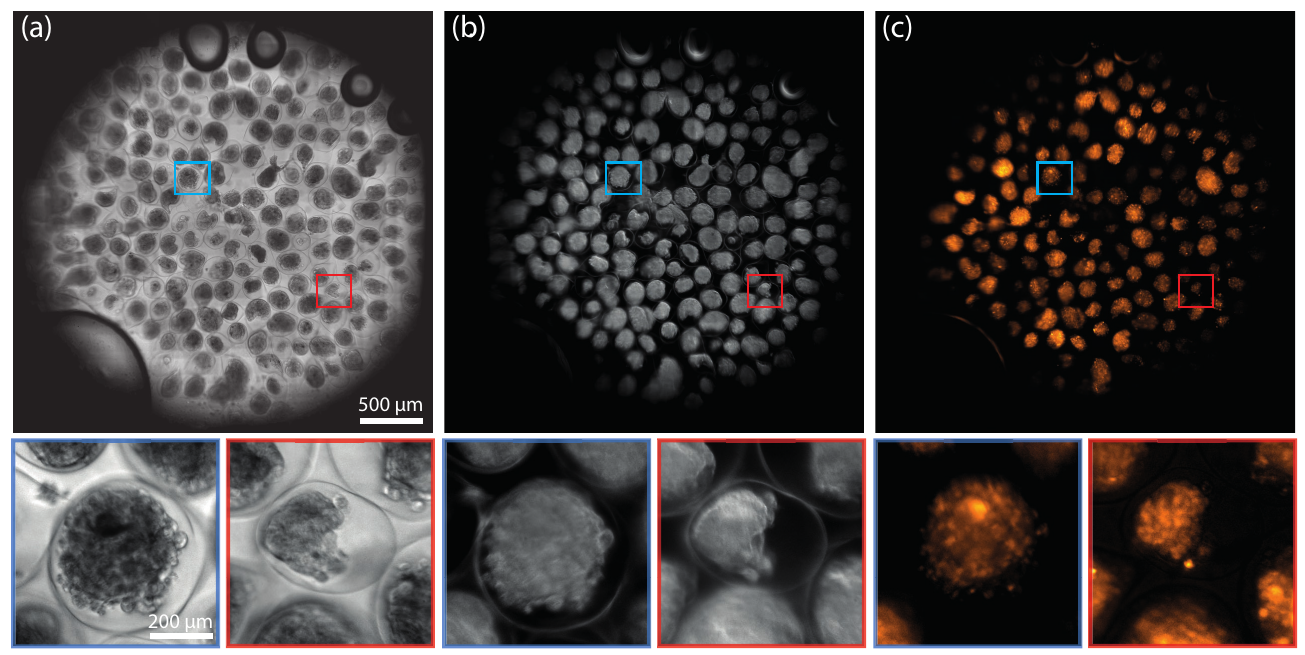}
\caption{Multi-modal acquisition of the entire field of view containing multi-cellular aggregates. (a-c) The same sample is imaged in bright field, phase contrast and epifluorescence modes. Insets represent close-up views of the region highlighted with a blue and red square in the large image.}\label{figure3}
\end{center}

\end{figure*}
\end{center}
Then, to cover the entire FOV of the MO at high lateral resolution, a sequence of 12 $\times$ 16 = 192 tilts is applied onto the motorized mirror and for each position of the sequence, the camera $\text{CMOS}_2$ captures an image. The whole sequence lasts about 8 seconds in bright-field mode. To combine the experimental data, we implemented a Fiji stitching plugin  \cite{preibisch2009globally} in Python through PyimageJ \cite{rueden2022pyimagej}. While the calibration step is initially time consuming, this procedure takes only a few seconds (<6 s) on a conventional computer once the coordinates are known. As a result, a 12 000 $\times$  12 500 pixels image is obtained, corresponding to a total of approximately 150 millions pixels. As seen on figure \ref{figure2}a, not all the pixels contains information and only a disk of diameter 11 000 pixels is filled, corresponding to 118 Mpixels containing information. The achieved value exceeds the theoretical SBP of 48 Mpixels, even though Nyquist-Shannon sampling is correctly satisfied. This discrepancy arises because the SBP depends on the ratio of the field number (FN), defined by the objective manufacturer, and the magnification Mag, which indicates the diffraction-limited performance area :
\begin{equation}
SBP=\frac{\pi^2 FN^2}{\lambda^2}. \left( \frac{NA^2}{Mag^2} \right)
\end{equation}
with $\lambda$ the wavelength. In practice, we collected data over a 3.3 mm FOV, larger than the expected 2.2 mm, though performance decreases at the edges, requiring careful characterization. To evaluate the optical performances of our apparatus, we first imaged sub-diffraction-sized beads on a glass slide in phase contrast mode. As seen on figure \ref{figure2}b,  a relatively homogeneous lateral resolution over a diameter of 2.2 mm is obtained which corresponds to the field number (FN) provided by the manufacturer. In this region, the performances are close to the theoretical resolution equals to 0.7 µm. At the edges of the FOV, coma aberrations degrade lateral resolution to 2-3 µm, making it non-diffraction limited but still usable. Additionally, the collection efficiency of the apparatus was estimated over the entire FOV and homogeneous performances were observed over more than 2 mm before a quick drop appears. These measurements show that our apparatus provides satisfactory optical performances over a 3 mm diameter FOV with a 10X microscope objective, and the detection unit introduces no additional aberrations since the scanning mechanism is positioned in the Fourier plane, where tilts are solely converted into shifts in the image plane.

Our approach is not limited to BF mode and its versatility can be demonstrated by imaging multi-cellular aggregates in phase contrast and epi-fluorescence modes under the same experimental conditions (see Figure \ref{figure3}). The imaging modalities offer complementary information; while the contrast in BF mode is mostly related to the absorption and scattering properties of the sample, phase contrast microscopy provides an increased contrast on transparent areas with refractive index differences. The BF mode offers a direct view of the size and shape of about 100  multi-cellular aggregates while the phase contrast mode highlights the interfaces such as the one between the alginate shell of the capsules and the surrounding medium. Finally, epifluorescence mode with an excitation at 488 nm provides a highly contrasted image with specific information that depends on the labeling strategy, here a nuclear marker. Note that in the latter case, exposure time had to be increased to 180 ms for each single image which turns the total acquisition time equals to approximately 30 seconds.

\begin{figure*}[ht!]
\begin{center}
\includegraphics[width=0.85\linewidth]{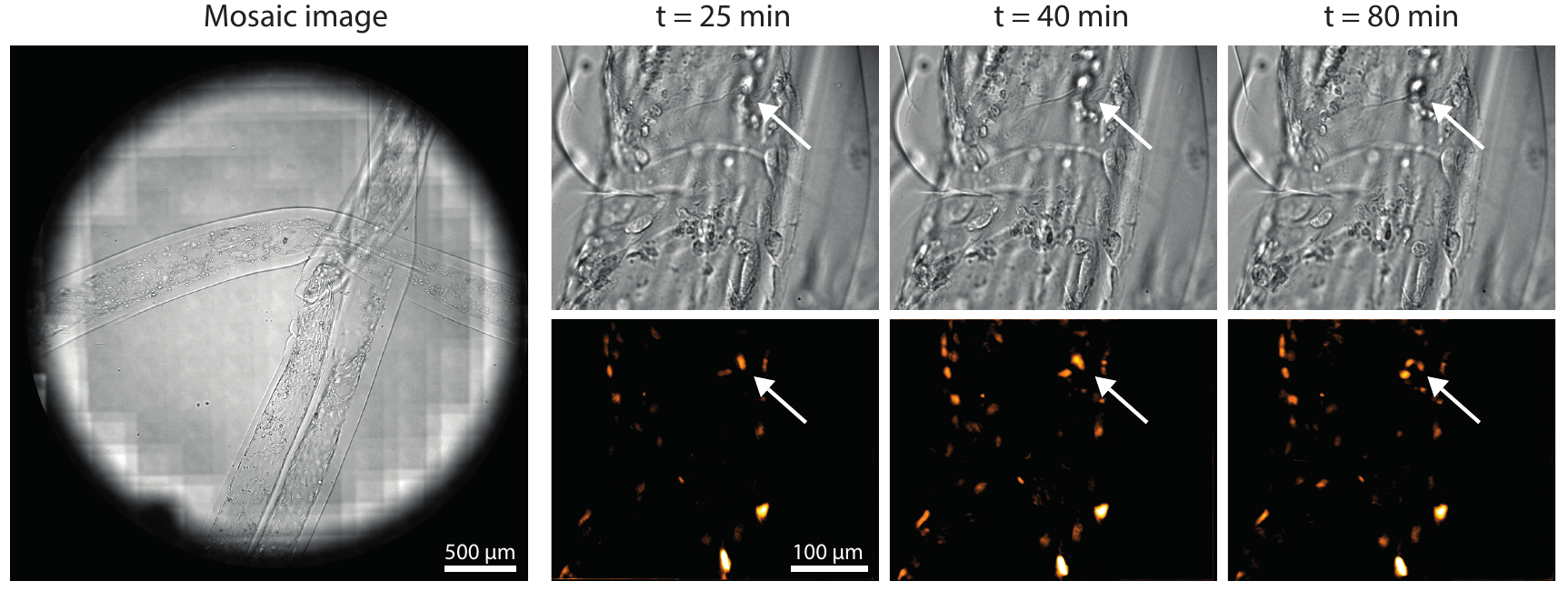}
\caption{Sparse ROI scanning. (a) A sequence of ROI is manually selected within the field of view. Scale bar 100 $\mu$m. (b) Only the selected ROI are captured with the selected total duration and time step between consecuvite acquisitions. Scale bar 100 $\mu$m. }
\label{figure4}
\end{center}
\end{figure*}
Capturing the entire FOV of the MO is often unnecessary and inefficient, as it generates excessive, mostly irrelevant data when only a few regions of interest are needed. Instead of applying a long sequence to homogeneously cover the FOV, the versatility of our detection unit allows to define a sequence that will target only several and discrete ROIs across the whole FOV.  This operation is particularly useful to image a sparse population of encapsulated cells placed in a \textit{Petri} dish or cells encapsulated inside a tubular alginate sheath (see Fig. \ref{figure4}). Depending on the nature of the encapsulated cells, these vesseloids can serve as \textit{in vitro} models of blood or lymphatic vessels \cite{andrique2019model}. For this example, 5 ROIs were manually selected, each of them being imaged every 4 minutes for a total duration of 3 hours. On a timelapse of a few hours, cell movements inside the alginate tube can be observed in both BF and epifluorescence modes, within a bio-engineered assembly of several millimeters. We were able to observe the temporal evolution of these cellular aggregates at high spatial resolution with experimental files for a total of 755 Mo only. This amount is very small compared to the 16 Go that would have been produced if the whole FOV had been captured.

In summary, we have developed a detection module compatible with any inverted commercial microscope that provides (1) a simultaneous observation of a large FOV at low-resolution and (2) a tunable smaller FOV at diffraction-limited resolution. Although the benefit of our approach is highly dependent on the performances of the MO and in particular its SBP, unlike previous work specific to a particular high-performance MO \cite{potsaid2005adaptive}, our approach is compatible with all conventional objectives. The main advantages of our system are that it is compact, fast, cost-effective, and user-friendly. First, our module fits on a single breadboard (25 x 30 $\text{cm}^2$) and can be easily installed on the output port of a microscope similarly to many other detection modules that already exist and provides a wide variety of modalities, from spinning disk microscopy \cite{cicero, aurox} to super-resolution techniques \cite{abbelight, abberior, confocal_nl} . Its simple and open design makes it easy to replace components such as the cameras, the beam-splitter or even to add new ones to expand the capabilities (tunable lenses, filters, \textit{etc}). Secondly, the speed of our system depends on the camera exposure time and the motorized mirror response time. As this response time is relatively small in our case ($\sim$ 3 ms), the total acquisition time is mainly limited by the exposure time of a single image multiplied by the number of images in the acquisition sequence. Yet, in the case of shorter exposure times, the total acquisition time of a sequence can be further reduced by replacing the motorized mirror by a faster mechanism such as galvanometric mirrors. Note that this solution is more costly, especially as large mirrors are needed. Third, we developed a simple graphical user interface to control both cameras and the motorized mirror using only USB connections. Information on how to update the code is provided on the GitHub page of our group \cite{github_bioflab}. Another advantage of our approach is that only one pre-calibration is needed to measure the positions of the images captured depending on the mirror positions. Once this is performed, the only post-processing step applied is the stitching algorithm to obtain an increased FOV.  All these elements mean that our system can be controlled by a computer with reasonable performances, no powerful processor or GPU is needed, which keeps our system user-friendly and cost-effective.

In this work, our system was designed with only two channels: a static low-resolution channel and a tunable high-resolution channel. However, a key advantage of our approach is its scalability to multiple channels, though this comes at the cost of distributing photons among them. This could enable several high-resolution tunable channels to capture simultaneously different locations, depths \cite{xiao2020high}, polarization states, or fluorescence channels. Such extensions are especially well-suited for BF modalities where the photon budget is relatively high. We can also envisage applying our general approach to other imaging modalities, in particular those that perform spatial multiplexing leading to a reduction in the FOV (quantitative phase imaging \cite{berto2017wavefront, bon2009quadriwave, antipa2018diffusercam} or light-field microscopy \cite{bazin2025increasing} for instance). In conclusion, the increase of the field of view, the simplicity, compactness and ease of use of our system should make it attractive for general biomedical research applications.

\textbf{Funding}

The authors acknowledge the financial support from the French National Agency for Research (ANR-22-CE42-0019), Idex Bordeaux (Research Program GPR Light), the Interdisciplinary and Exploratory Research grant from Bordeaux University. This work has benefited from a government grant managed by the Agence Nationale de la Recherche under the France 2030 program, reference ANR-24-EXME-0002. \\

\textbf{Acknowledgments}

We would like to thank all the BiOf team for fruitful discussions and especially Camille Douillet, Elsa Mazari-Arrighi, Aurélien Richard and Adeline Boyreau for providing biological samples. \\

\textbf{Disclosures}

A.B., Gaëlle Recher and Pierre Nassoy are inventors on a patent related to this work (no. WO2023036946A1). All authors declare that they have no other competing interests.\\

\textbf{Data availability} 

Data underlying the results presented in this paper can be obtained from the authors upon reasonable request. \\

%%%%%%%%%%%%%%%%%%%%%%% References %%%%%%%%%%%%%%%%%%%%%%%%%

%%%%%%%%%% If using BibTeX:
\bibliography{Letter_draft}

\begin{thebibliography}{24}%
\makeatletter
\providecommand \@ifxundefined [1]{%
 \@ifx{#1\undefined}
}%
\providecommand \@ifnum [1]{%
 \ifnum #1\expandafter \@firstoftwo
 \else \expandafter \@secondoftwo
 \fi
}%
\providecommand \@ifx [1]{%
 \ifx #1\expandafter \@firstoftwo
 \else \expandafter \@secondoftwo
 \fi
}%
\providecommand \natexlab [1]{#1}%
\providecommand \enquote  [1]{``#1''}%
\providecommand \bibnamefont  [1]{#1}%
\providecommand \bibfnamefont [1]{#1}%
\providecommand \citenamefont [1]{#1}%
\providecommand \href@noop [0]{\@secondoftwo}%
\providecommand \href [0]{\begingroup \@sanitize@url \@href}%
\providecommand \@href[1]{\@@startlink{#1}\@@href}%
\providecommand \@@href[1]{\endgroup#1\@@endlink}%
\providecommand \@sanitize@url [0]{\catcode `\\12\catcode `\$12\catcode `\&12\catcode `\#12\catcode `\^12\catcode `\_12\catcode `\%12\relax}%
\providecommand \@@startlink[1]{}%
\providecommand \@@endlink[0]{}%
\providecommand \url  [0]{\begingroup\@sanitize@url \@url }%
\providecommand \@url [1]{\endgroup\@href {#1}{\urlprefix }}%
\providecommand \urlprefix  [0]{URL }%
\providecommand \Eprint [0]{\href }%
\providecommand \doibase [0]{http://dx.doi.org/}%
\providecommand \selectlanguage [0]{\@gobble}%
\providecommand \bibinfo  [0]{\@secondoftwo}%
\providecommand \bibfield  [0]{\@secondoftwo}%
\providecommand \translation [1]{[#1]}%
\providecommand \BibitemOpen [0]{}%
\providecommand \bibitemStop [0]{}%
\providecommand \bibitemNoStop [0]{.\EOS\space}%
\providecommand \EOS [0]{\spacefactor3000\relax}%
\providecommand \BibitemShut  [1]{\csname bibitem#1\endcsname}%
\let\auto@bib@innerbib\@empty
%</preamble>
\bibitem [{\citenamefont {Yang}\ \emph {et~al.}(2022)\citenamefont {Yang}, \citenamefont {Lange}, \citenamefont {Millett-Sikking}, \citenamefont {Zhao}, \citenamefont {Bragantini}, \citenamefont {VijayKumar}, \citenamefont {Kamb}, \citenamefont {G{\'o}mez-Sj{\"o}berg}, \citenamefont {Solak}, \citenamefont {Wang} \emph {et~al.}}]{yang2022daxi}%
  \BibitemOpen
  \bibfield  {author} {\bibinfo {author} {\bibfnamefont {B.}~\bibnamefont {Yang}}, \bibinfo {author} {\bibfnamefont {M.}~\bibnamefont {Lange}}, \bibinfo {author} {\bibfnamefont {A.}~\bibnamefont {Millett-Sikking}}, \bibinfo {author} {\bibfnamefont {X.}~\bibnamefont {Zhao}}, \bibinfo {author} {\bibfnamefont {J.}~\bibnamefont {Bragantini}}, \bibinfo {author} {\bibfnamefont {S.}~\bibnamefont {VijayKumar}}, \bibinfo {author} {\bibfnamefont {M.}~\bibnamefont {Kamb}}, \bibinfo {author} {\bibfnamefont {R.}~\bibnamefont {G{\'o}mez-Sj{\"o}berg}}, \bibinfo {author} {\bibfnamefont {A.~C.}\ \bibnamefont {Solak}}, \bibinfo {author} {\bibfnamefont {W.}~\bibnamefont {Wang}},  \emph {et~al.},\ }\href@noop {} {\bibfield  {journal} {\bibinfo  {journal} {Nature methods}\ }\textbf {\bibinfo {volume} {19}},\ \bibinfo {pages} {461} (\bibinfo {year} {2022})}\BibitemShut {NoStop}%
\bibitem [{\citenamefont {Glaser}\ \emph {et~al.}(2022)\citenamefont {Glaser}, \citenamefont {Bishop}, \citenamefont {Barner}, \citenamefont {Susaki}, \citenamefont {Kubota}, \citenamefont {Gao}, \citenamefont {Serafin}, \citenamefont {Balaram}, \citenamefont {Turschak}, \citenamefont {Nicovich} \emph {et~al.}}]{glaser2022hybrid}%
  \BibitemOpen
  \bibfield  {author} {\bibinfo {author} {\bibfnamefont {A.~K.}\ \bibnamefont {Glaser}}, \bibinfo {author} {\bibfnamefont {K.~W.}\ \bibnamefont {Bishop}}, \bibinfo {author} {\bibfnamefont {L.~A.}\ \bibnamefont {Barner}}, \bibinfo {author} {\bibfnamefont {E.~A.}\ \bibnamefont {Susaki}}, \bibinfo {author} {\bibfnamefont {S.~I.}\ \bibnamefont {Kubota}}, \bibinfo {author} {\bibfnamefont {G.}~\bibnamefont {Gao}}, \bibinfo {author} {\bibfnamefont {R.~B.}\ \bibnamefont {Serafin}}, \bibinfo {author} {\bibfnamefont {P.}~\bibnamefont {Balaram}}, \bibinfo {author} {\bibfnamefont {E.}~\bibnamefont {Turschak}}, \bibinfo {author} {\bibfnamefont {P.~R.}\ \bibnamefont {Nicovich}},  \emph {et~al.},\ }\href@noop {} {\bibfield  {journal} {\bibinfo  {journal} {Nature methods}\ }\textbf {\bibinfo {volume} {19}},\ \bibinfo {pages} {613} (\bibinfo {year} {2022})}\BibitemShut {NoStop}%
\bibitem [{\citenamefont {Lohmann}\ \emph {et~al.}(1996)\citenamefont {Lohmann}, \citenamefont {Dorsch}, \citenamefont {Mendlovic}, \citenamefont {Zalevsky},\ and\ \citenamefont {Ferreira}}]{lohmann1996space}%
  \BibitemOpen
  \bibfield  {author} {\bibinfo {author} {\bibfnamefont {A.~W.}\ \bibnamefont {Lohmann}}, \bibinfo {author} {\bibfnamefont {R.~G.}\ \bibnamefont {Dorsch}}, \bibinfo {author} {\bibfnamefont {D.}~\bibnamefont {Mendlovic}}, \bibinfo {author} {\bibfnamefont {Z.}~\bibnamefont {Zalevsky}}, \ and\ \bibinfo {author} {\bibfnamefont {C.}~\bibnamefont {Ferreira}},\ }\href@noop {} {\bibfield  {journal} {\bibinfo  {journal} {Journal of the Optical Society of America A}\ }\textbf {\bibinfo {volume} {13}},\ \bibinfo {pages} {470} (\bibinfo {year} {1996})}\BibitemShut {NoStop}%
\bibitem [{\citenamefont {Park}\ \emph {et~al.}(2021)\citenamefont {Park}, \citenamefont {Brady}, \citenamefont {Zheng}, \citenamefont {Tian},\ and\ \citenamefont {Gao}}]{park2021review}%
  \BibitemOpen
  \bibfield  {author} {\bibinfo {author} {\bibfnamefont {J.}~\bibnamefont {Park}}, \bibinfo {author} {\bibfnamefont {D.~J.}\ \bibnamefont {Brady}}, \bibinfo {author} {\bibfnamefont {G.}~\bibnamefont {Zheng}}, \bibinfo {author} {\bibfnamefont {L.}~\bibnamefont {Tian}}, \ and\ \bibinfo {author} {\bibfnamefont {L.}~\bibnamefont {Gao}},\ }\href@noop {} {\bibfield  {journal} {\bibinfo  {journal} {Advanced Photonics}\ }\textbf {\bibinfo {volume} {3}},\ \bibinfo {pages} {044001} (\bibinfo {year} {2021})}\BibitemShut {NoStop}%
\bibitem [{\citenamefont {Zheng}\ \emph {et~al.}(2013)\citenamefont {Zheng}, \citenamefont {Horstmeyer},\ and\ \citenamefont {Yang}}]{zheng2013wide}%
  \BibitemOpen
  \bibfield  {author} {\bibinfo {author} {\bibfnamefont {G.}~\bibnamefont {Zheng}}, \bibinfo {author} {\bibfnamefont {R.}~\bibnamefont {Horstmeyer}}, \ and\ \bibinfo {author} {\bibfnamefont {C.}~\bibnamefont {Yang}},\ }\href@noop {} {\bibfield  {journal} {\bibinfo  {journal} {Nature photonics}\ }\textbf {\bibinfo {volume} {7}},\ \bibinfo {pages} {739} (\bibinfo {year} {2013})}\BibitemShut {NoStop}%
\bibitem [{\citenamefont {Brady}\ \emph {et~al.}(2012)\citenamefont {Brady}, \citenamefont {Gehm}, \citenamefont {Stack}, \citenamefont {Marks}, \citenamefont {Kittle}, \citenamefont {Golish}, \citenamefont {Vera},\ and\ \citenamefont {Feller}}]{brady2012multiscale}%
  \BibitemOpen
  \bibfield  {author} {\bibinfo {author} {\bibfnamefont {D.~J.}\ \bibnamefont {Brady}}, \bibinfo {author} {\bibfnamefont {M.~E.}\ \bibnamefont {Gehm}}, \bibinfo {author} {\bibfnamefont {R.~A.}\ \bibnamefont {Stack}}, \bibinfo {author} {\bibfnamefont {D.~L.}\ \bibnamefont {Marks}}, \bibinfo {author} {\bibfnamefont {D.~S.}\ \bibnamefont {Kittle}}, \bibinfo {author} {\bibfnamefont {D.~R.}\ \bibnamefont {Golish}}, \bibinfo {author} {\bibfnamefont {E.}~\bibnamefont {Vera}}, \ and\ \bibinfo {author} {\bibfnamefont {S.~D.}\ \bibnamefont {Feller}},\ }\href@noop {} {\bibfield  {journal} {\bibinfo  {journal} {Nature}\ }\textbf {\bibinfo {volume} {486}},\ \bibinfo {pages} {386} (\bibinfo {year} {2012})}\BibitemShut {NoStop}%
\bibitem [{\citenamefont {Gustafsson}(2000)}]{gustafsson2000surpassing}%
  \BibitemOpen
  \bibfield  {author} {\bibinfo {author} {\bibfnamefont {M.~G.}\ \bibnamefont {Gustafsson}},\ }\href@noop {} {\bibfield  {journal} {\bibinfo  {journal} {Journal of microscopy}\ }\textbf {\bibinfo {volume} {198}},\ \bibinfo {pages} {82} (\bibinfo {year} {2000})}\BibitemShut {NoStop}%
\bibitem [{\citenamefont {Alessandri}\ \emph {et~al.}(2013)\citenamefont {Alessandri}, \citenamefont {Sarangi}, \citenamefont {Gurchenkov}, \citenamefont {Sinha}, \citenamefont {Kie{\ss}ling}, \citenamefont {Fetler}, \citenamefont {Rico}, \citenamefont {Scheuring}, \citenamefont {Lamaze}, \citenamefont {Simon} \emph {et~al.}}]{alessandri2013cellular}%
  \BibitemOpen
  \bibfield  {author} {\bibinfo {author} {\bibfnamefont {K.}~\bibnamefont {Alessandri}}, \bibinfo {author} {\bibfnamefont {B.~R.}\ \bibnamefont {Sarangi}}, \bibinfo {author} {\bibfnamefont {V.~V.}\ \bibnamefont {Gurchenkov}}, \bibinfo {author} {\bibfnamefont {B.}~\bibnamefont {Sinha}}, \bibinfo {author} {\bibfnamefont {T.~R.}\ \bibnamefont {Kie{\ss}ling}}, \bibinfo {author} {\bibfnamefont {L.}~\bibnamefont {Fetler}}, \bibinfo {author} {\bibfnamefont {F.}~\bibnamefont {Rico}}, \bibinfo {author} {\bibfnamefont {S.}~\bibnamefont {Scheuring}}, \bibinfo {author} {\bibfnamefont {C.}~\bibnamefont {Lamaze}}, \bibinfo {author} {\bibfnamefont {A.}~\bibnamefont {Simon}},  \emph {et~al.},\ }\href@noop {} {\bibfield  {journal} {\bibinfo  {journal} {Proceedings of the National Academy of Sciences}\ }\textbf {\bibinfo {volume} {110}},\ \bibinfo {pages} {14843} (\bibinfo {year} {2013})}\BibitemShut {NoStop}%
\bibitem [{\citenamefont {Cohen}\ \emph {et~al.}(2023)\citenamefont {Cohen}, \citenamefont {Luquet}, \citenamefont {Pletenka}, \citenamefont {Leonard}, \citenamefont {Warter}, \citenamefont {Gurchenkov}, \citenamefont {Carrere}, \citenamefont {Rieu}, \citenamefont {Hardouin}, \citenamefont {Moncaubeig} \emph {et~al.}}]{cohen2023engineering}%
  \BibitemOpen
  \bibfield  {author} {\bibinfo {author} {\bibfnamefont {P.~J.}\ \bibnamefont {Cohen}}, \bibinfo {author} {\bibfnamefont {E.}~\bibnamefont {Luquet}}, \bibinfo {author} {\bibfnamefont {J.}~\bibnamefont {Pletenka}}, \bibinfo {author} {\bibfnamefont {A.}~\bibnamefont {Leonard}}, \bibinfo {author} {\bibfnamefont {E.}~\bibnamefont {Warter}}, \bibinfo {author} {\bibfnamefont {B.}~\bibnamefont {Gurchenkov}}, \bibinfo {author} {\bibfnamefont {J.}~\bibnamefont {Carrere}}, \bibinfo {author} {\bibfnamefont {C.}~\bibnamefont {Rieu}}, \bibinfo {author} {\bibfnamefont {J.}~\bibnamefont {Hardouin}}, \bibinfo {author} {\bibfnamefont {F.}~\bibnamefont {Moncaubeig}},  \emph {et~al.},\ }\href@noop {} {\bibfield  {journal} {\bibinfo  {journal} {Biomaterials}\ }\textbf {\bibinfo {volume} {295}},\ \bibinfo {pages} {122033} (\bibinfo {year} {2023})}\BibitemShut {NoStop}%
\bibitem [{git()}]{github_bioflab}%
  \BibitemOpen
  \href@noop {} {}\bibinfo {howpublished} {\url{https://github.com/BiOflab/Universcope}}\BibitemShut {NoStop}%
\bibitem [{\citenamefont {Preibisch}\ \emph {et~al.}(2009)\citenamefont {Preibisch}, \citenamefont {Saalfeld},\ and\ \citenamefont {Tomancak}}]{preibisch2009globally}%
  \BibitemOpen
  \bibfield  {author} {\bibinfo {author} {\bibfnamefont {S.}~\bibnamefont {Preibisch}}, \bibinfo {author} {\bibfnamefont {S.}~\bibnamefont {Saalfeld}}, \ and\ \bibinfo {author} {\bibfnamefont {P.}~\bibnamefont {Tomancak}},\ }\href@noop {} {\bibfield  {journal} {\bibinfo  {journal} {Bioinformatics}\ }\textbf {\bibinfo {volume} {25}},\ \bibinfo {pages} {1463} (\bibinfo {year} {2009})}\BibitemShut {NoStop}%
\bibitem [{\citenamefont {Rueden}\ \emph {et~al.}(2022)\citenamefont {Rueden}, \citenamefont {Hiner}, \citenamefont {Evans~III}, \citenamefont {Pinkert}, \citenamefont {Lucas}, \citenamefont {Carpenter}, \citenamefont {Cimini},\ and\ \citenamefont {Eliceiri}}]{rueden2022pyimagej}%
  \BibitemOpen
  \bibfield  {author} {\bibinfo {author} {\bibfnamefont {C.~T.}\ \bibnamefont {Rueden}}, \bibinfo {author} {\bibfnamefont {M.~C.}\ \bibnamefont {Hiner}}, \bibinfo {author} {\bibfnamefont {E.~L.}\ \bibnamefont {Evans~III}}, \bibinfo {author} {\bibfnamefont {M.~A.}\ \bibnamefont {Pinkert}}, \bibinfo {author} {\bibfnamefont {A.~M.}\ \bibnamefont {Lucas}}, \bibinfo {author} {\bibfnamefont {A.~E.}\ \bibnamefont {Carpenter}}, \bibinfo {author} {\bibfnamefont {B.~A.}\ \bibnamefont {Cimini}}, \ and\ \bibinfo {author} {\bibfnamefont {K.~W.}\ \bibnamefont {Eliceiri}},\ }\href@noop {} {\bibfield  {journal} {\bibinfo  {journal} {Nature methods}\ }\textbf {\bibinfo {volume} {19}},\ \bibinfo {pages} {1326} (\bibinfo {year} {2022})}\BibitemShut {NoStop}%
\bibitem [{\citenamefont {Andrique}\ \emph {et~al.}(2019)\citenamefont {Andrique}, \citenamefont {Recher}, \citenamefont {Alessandri}, \citenamefont {Pujol}, \citenamefont {Feyeux}, \citenamefont {Bon}, \citenamefont {Cognet}, \citenamefont {Nassoy},\ and\ \citenamefont {Bikfalvi}}]{andrique2019model}%
  \BibitemOpen
  \bibfield  {author} {\bibinfo {author} {\bibfnamefont {L.}~\bibnamefont {Andrique}}, \bibinfo {author} {\bibfnamefont {G.}~\bibnamefont {Recher}}, \bibinfo {author} {\bibfnamefont {K.}~\bibnamefont {Alessandri}}, \bibinfo {author} {\bibfnamefont {N.}~\bibnamefont {Pujol}}, \bibinfo {author} {\bibfnamefont {M.}~\bibnamefont {Feyeux}}, \bibinfo {author} {\bibfnamefont {P.}~\bibnamefont {Bon}}, \bibinfo {author} {\bibfnamefont {L.}~\bibnamefont {Cognet}}, \bibinfo {author} {\bibfnamefont {P.}~\bibnamefont {Nassoy}}, \ and\ \bibinfo {author} {\bibfnamefont {A.}~\bibnamefont {Bikfalvi}},\ }\href@noop {} {\bibfield  {journal} {\bibinfo  {journal} {Science advances}\ }\textbf {\bibinfo {volume} {5}},\ \bibinfo {pages} {eaau6562} (\bibinfo {year} {2019})}\BibitemShut {NoStop}%
\bibitem [{\citenamefont {Potsaid}\ \emph {et~al.}(2005)\citenamefont {Potsaid}, \citenamefont {Bellouard},\ and\ \citenamefont {Wen}}]{potsaid2005adaptive}%
  \BibitemOpen
  \bibfield  {author} {\bibinfo {author} {\bibfnamefont {B.}~\bibnamefont {Potsaid}}, \bibinfo {author} {\bibfnamefont {Y.}~\bibnamefont {Bellouard}}, \ and\ \bibinfo {author} {\bibfnamefont {J.~T.}\ \bibnamefont {Wen}},\ }\href@noop {} {\bibfield  {journal} {\bibinfo  {journal} {Optics express}\ }\textbf {\bibinfo {volume} {13}},\ \bibinfo {pages} {6504} (\bibinfo {year} {2005})}\BibitemShut {NoStop}%
\bibitem [{cic()}]{cicero}%
  \BibitemOpen
  \href@noop {} {}\bibinfo {howpublished} {\url{https://crestoptics.com/cicero/ }}\BibitemShut {NoStop}%
\bibitem [{aur()}]{aurox}%
  \BibitemOpen
  \href@noop {} {}\bibinfo {howpublished} {\url{https://www.aurox.co.uk/clarity.php }}\BibitemShut {NoStop}%
\bibitem [{abb({\natexlab{a}})}]{abbelight}%
  \BibitemOpen
  \href@noop {} {}\bibinfo {howpublished} {\url{https://www.abbelight.com}} ({\natexlab{a}})\BibitemShut {NoStop}%
\bibitem [{abb({\natexlab{b}})}]{abberior}%
  \BibitemOpen
  \href@noop {} {}\bibinfo {howpublished} {\url{https://abberior.rocks/superresolution-confocal-systems/modules/}} ({\natexlab{b}})\BibitemShut {NoStop}%
\bibitem [{con()}]{confocal_nl}%
  \BibitemOpen
  \href@noop {} {}\bibinfo {howpublished} {\url{https://www.confocal.nl }}\BibitemShut {NoStop}%
\bibitem [{\citenamefont {Xiao}\ \emph {et~al.}(2020)\citenamefont {Xiao}, \citenamefont {Gritton}, \citenamefont {Tseng}, \citenamefont {Zemel}, \citenamefont {Han},\ and\ \citenamefont {Mertz}}]{xiao2020high}%
  \BibitemOpen
  \bibfield  {author} {\bibinfo {author} {\bibfnamefont {S.}~\bibnamefont {Xiao}}, \bibinfo {author} {\bibfnamefont {H.}~\bibnamefont {Gritton}}, \bibinfo {author} {\bibfnamefont {H.-a.}\ \bibnamefont {Tseng}}, \bibinfo {author} {\bibfnamefont {D.}~\bibnamefont {Zemel}}, \bibinfo {author} {\bibfnamefont {X.}~\bibnamefont {Han}}, \ and\ \bibinfo {author} {\bibfnamefont {J.}~\bibnamefont {Mertz}},\ }\href@noop {} {\bibfield  {journal} {\bibinfo  {journal} {Optica}\ }\textbf {\bibinfo {volume} {7}},\ \bibinfo {pages} {1477} (\bibinfo {year} {2020})}\BibitemShut {NoStop}%
\bibitem [{\citenamefont {Berto}\ \emph {et~al.}(2017)\citenamefont {Berto}, \citenamefont {Rigneault},\ and\ \citenamefont {Guillon}}]{berto2017wavefront}%
  \BibitemOpen
  \bibfield  {author} {\bibinfo {author} {\bibfnamefont {P.}~\bibnamefont {Berto}}, \bibinfo {author} {\bibfnamefont {H.}~\bibnamefont {Rigneault}}, \ and\ \bibinfo {author} {\bibfnamefont {M.}~\bibnamefont {Guillon}},\ }\href@noop {} {\bibfield  {journal} {\bibinfo  {journal} {Optics Letters}\ }\textbf {\bibinfo {volume} {42}},\ \bibinfo {pages} {5117} (\bibinfo {year} {2017})}\BibitemShut {NoStop}%
\bibitem [{\citenamefont {Bon}\ \emph {et~al.}(2009)\citenamefont {Bon}, \citenamefont {Maucort}, \citenamefont {Wattellier},\ and\ \citenamefont {Monneret}}]{bon2009quadriwave}%
  \BibitemOpen
  \bibfield  {author} {\bibinfo {author} {\bibfnamefont {P.}~\bibnamefont {Bon}}, \bibinfo {author} {\bibfnamefont {G.}~\bibnamefont {Maucort}}, \bibinfo {author} {\bibfnamefont {B.}~\bibnamefont {Wattellier}}, \ and\ \bibinfo {author} {\bibfnamefont {S.}~\bibnamefont {Monneret}},\ }\href@noop {} {\bibfield  {journal} {\bibinfo  {journal} {Optics express}\ }\textbf {\bibinfo {volume} {17}},\ \bibinfo {pages} {13080} (\bibinfo {year} {2009})}\BibitemShut {NoStop}%
\bibitem [{\citenamefont {Antipa}\ \emph {et~al.}(2018)\citenamefont {Antipa}, \citenamefont {Kuo}, \citenamefont {Heckel}, \citenamefont {Mildenhall}, \citenamefont {Bostan}, \citenamefont {Ng},\ and\ \citenamefont {Waller}}]{antipa2018diffusercam}%
  \BibitemOpen
  \bibfield  {author} {\bibinfo {author} {\bibfnamefont {N.}~\bibnamefont {Antipa}}, \bibinfo {author} {\bibfnamefont {G.}~\bibnamefont {Kuo}}, \bibinfo {author} {\bibfnamefont {R.}~\bibnamefont {Heckel}}, \bibinfo {author} {\bibfnamefont {B.}~\bibnamefont {Mildenhall}}, \bibinfo {author} {\bibfnamefont {E.}~\bibnamefont {Bostan}}, \bibinfo {author} {\bibfnamefont {R.}~\bibnamefont {Ng}}, \ and\ \bibinfo {author} {\bibfnamefont {L.}~\bibnamefont {Waller}},\ }\href@noop {} {\bibfield  {journal} {\bibinfo  {journal} {Optica}\ }\textbf {\bibinfo {volume} {5}},\ \bibinfo {pages} {1} (\bibinfo {year} {2018})}\BibitemShut {NoStop}%
\bibitem [{\citenamefont {Bazin}\ and\ \citenamefont {Badon}(2025)}]{bazin2025increasing}%
  \BibitemOpen
  \bibfield  {author} {\bibinfo {author} {\bibfnamefont {A.}~\bibnamefont {Bazin}}\ and\ \bibinfo {author} {\bibfnamefont {A.}~\bibnamefont {Badon}},\ }\href@noop {} {\bibfield  {journal} {\bibinfo  {journal} {Biomedical Optics Express}\ }\textbf {\bibinfo {volume} {16}},\ \bibinfo {pages} {1062} (\bibinfo {year} {2025})}\BibitemShut {NoStop}%
\end{thebibliography}%

\end{document}